# Physics to System-level Modeling of Silicon-organic-hybrid Nanophotonic Devices

**Maryam Moridsadat[1], Marcus Tamura[1], Lukas Chrostowski[2], Sudip Shekhar[2], Bhavin J. Shastri[1]**

*Corresponding Authors: Bhavin J. Shastri (Primary), Maryam Moridsadat*

[1] Center of Nanophotonics, Department of Physics, Engineering Physics & Astronomy, Queen's University, Kingston, ON K7L 3N6, Canada
[2] Department of Electrical and Computer Engineering, University of British Columbia, Vancouver, BC V6T 1Z4, Canada

## Abstract

The continuous growth in data volume has sparked interest in silicon-organic-hybrid (SOH) nanophotonic devices integrated into silicon photonic integrated circuits (PICs). SOH devices offer improved speed and energy efficiency compared to silicon photonics devices. However, a comprehensive and accurate modeling methodology of SOH devices, such as modulators corroborating experimental results, is lacking. While some preliminary modeling approaches for SOH devices exist, their reliance on theoretical and numerical methodologies, along with a lack of compatibility with electronic design automation (EDA), hinders their seamless and rapid integration with silicon PICs. Here, we develop a phenomenological, building-block-based SOH PICs simulation methodology that spans from the physics to the system level, offering high accuracy, comprehensiveness, and EDA-style compatibility. Our model is also readily integrable and scalable, lending itself to the design of large-scale silicon PICs. Our proposed modeling methodology is agnostic and compatible with any photonics-electronics co-simulation software. We validate this methodology by comparing the characteristics of experimentally demonstrated SOH microring modulators (MRMs) and Mach Zehnder modulators (MZMs) with those obtained through simulation, demonstrating its ability to model various modulator topologies. We also show our methodology's ease and speed in modeling large-scale systems. As an illustrative example, we use our methodology to design and study a 3-channel SOH MRM-based wavelength-division (de)multiplexer, a widely used component in various applications, including neuromorphic computing, data center interconnects, communications, sensing, and switching networks. Our modeling approach is also compatible with other materials exhibiting the Pockels and Kerr effects. To our knowledge, this represents the first comprehensive physics-to-system-level EDA-compatible simulation methodology for SOH modulators.

## 1. Introduction

Silicon photonics technology compatible with complementary Metal-Oxide Semiconductor (CMOS) has facilitated large-scale integration of photonic components enabling various applications, including data transmission [1,2], sensing [3], quantum computing [4,5], and neuromorphic computing [6-8]. The practical realization of photonic integrated circuits (PICs) from design to fabrication requires different simulation platforms for simultaneously designing and simulating photonic and electronic devices as part of a single system [9,10]. Inspired by electronic design automation (EDA) [11-18], Chrostowski et al. [11,13,14], Pond et al. [12,15], and Wang et al. [17] have introduced a well-established EDA-style automated full-flow design methodology for silicon photonics circuits that has enabled the creation of PICs from design to fabrication.

Relentless data volume growth worldwide has sparked a strong interest in hybrid systems combining emerging materials with silicon in PICs [19-21], including graphene [22-24], ITO [25-27], LiNbO$_3$ [28-31], and polymers [19,32-35]. Polymers are known as one of the most promising groups owing to their high-speed, low-cost, and low-power consumption [35-39]. Several studies have examined silicon-organic hybrid (SOH) structures, including microring resonators (MRRs) [40-43] and Mach-Zehnder interferometers (MZIs) [34,44-47]. Incorporating electro-optic (EO) SOH modulators in silicon PICs holds the promise of keeping pace with the ever-increasing data rate. This necessitates the development of a novel and comprehensive simulation methodology for SOH modulators compatible with well-established, fully automated EDA-style silicon PIC design methodologies [11-18]. However, only some approaches explore the theoretical, numerical, and mix of theoretical and numerical modeling of polymer-based modulators. Recently, Tibaldi et al. [48] introduced a physics-level to system-level theoretical modeling approach of SOH MZMs using a series of formulas. A theoretical modeling methodology that relies on complex sets of formulations considering all physics-level to device-level parameters impacting modulators performance remains a challenge for photonic designers in designing SOH modulators [49]. Tibaldi et al. [48] also ignored the high experimental loss of slotted waveguide SOH modulators, resulting in an incorrect extinction ratio and quality factor [50]. Some studies have also combined the software-based 2D simulation and theoretical models to explain the EO active response observed in experiments [34,51,52]. These works have only studied the EO response of the SOH modulators and have yet to discuss the passive optical simulation (e.g., quality factor, extinction ratio, free spectral range (FSR), and resonance wavelength), all of which are required in designing SOH devices. Furthermore, simplified theoretical calculations can be found in the literature to estimate the EO behavior of SOH modulators [44,53]. In these studies, a spatially constant electrical electric field (E-field) value in the slot region and negligible electrical E-field outside of the slot have been considered, which results in an inaccurate EO active simulation of SOH modulators with passive optical simulation overlooked [44,53]. All the studied modeling approaches are also incompatible with automated device-to-system EDA-style design workflow from simulation to fabrication [11-18]. Currently, there is no comprehensive, accurate and efficient EDA-style compatible software-based modeling approach to design and simulate SOH modulators.

Here, we develop a thorough physics to system-level building block-based phenomenological simulation methodology for SOH modulators. Our approach is built upon standard simulation software and is compatible with EDA-style silicon PICs modeling strategies. Our model accounts for the experimental loss and the nonuniform electrical E-field in the core and cladding of the slotted waveguide to accurately predict the experimental results. The proposed simulation methodology can be implemented with any multi-physics simulation software supporting optical and electrical simulations. We validate the proposed simulation methodology CHARGE ( on optical and electrical physics-based and circuit-level solvers of the well-known Ansys/Lumerical [54] commercial software, including MODE (optical waveguide and coupler solver), FDTD (photonics components simulator), CHARGE (charge transport simulator), and INTERCONNECT (system simulator) packages. These packages enable the photonic-electronic co-simulations of optoelectronic SOH modulators in a single environment. We validate our simulation methodology by comparing the device performance of experimentally demonstrated SOH MRM [40] and MZM [44] with those obtained from the simulations while showing the ability of our simulation methodology to model a broad class of EO SOH modulator topologies. Our building block-based modeling methodology enables the fast and easy simulation of large-scale systems , which paves the way for future applications of SOH-based structures in PICs. As a use case example, we design, model, and analyze a three-channel SOH MRM-based wave division (WDM) (de)multiplexer which is widely used in different applications like neuromorphic computing, datacenters interconnects, communications, sensing and switching networks. The simulation methodology also enables PDK library integration. To the best of our knowledge, this is the first comprehensive physics-based to system-level EDA-compatible simulation methodology of SOH structures.

## 2. Simulation Methodology

### 2.1. Device-level passive optical simulations

One approach for simulating photonic devices is to model the entire photonic structure as a single 3D optical system—a so-called all-in-one (AiO) strategy [48,55]. The intensive computational burden of AiO 3D simulation makes the modeling prohibitive for structures with lengths greater than $10\mu m$ [48,55]. MZMs inherently have a large footprint due to their non-resonant behavior. While MRMs have a smaller footprint compared to MZMs, the slotted waveguides SOH MRMs are relatively larger than their silicon counterparts to mitigate the high slotted waveguides bending loss. Hence, the AiO 3D finite difference time domain (FDTD) simulation is impractical, for example, for simulating SOH MRRs with radii of 40 $\mu m$ [41] and 60 $\mu m$ [40] and SOH MZMs lengths of 0.5mm [39] and 1.1mm [44].Our proposed simulation methodology for SOH devices has been adapted based on the building block method instead of the AiO approach. In our approach, the structure is divided into subcomponents, and the device-level physical simulations are focused on individual building blocks to find their physical characteristics as a scattering matrix. Then, the whole system is modeled by combining the characterized sub-components with their associated scattering matrices [48,56-58]. The building block approach also prevents repeated time-consuming 3D optical simulations at each voltage for SOH modulators. While our methodology is agnostic to the simulation platform, we use the Ansys/Lumerical photonic simulator with the MODE, FDTD, CHARGE, and INTERCONNECT packages. Fig. 1 represents the graphical overview of the developed simulation workflow. Other simulation software can be used in place of Ansys/Lumerical software such as MEEP, an open-source photonics component simulator written in Python. Additionally, Cadence Virtuoso can utilize Verilog-A models to simulate circuit level photonic circuits [59,60]. We use the Ansys/Lumerical simulators in this work since importing the data within the Ansys/Lumerical workflow is streamlined.

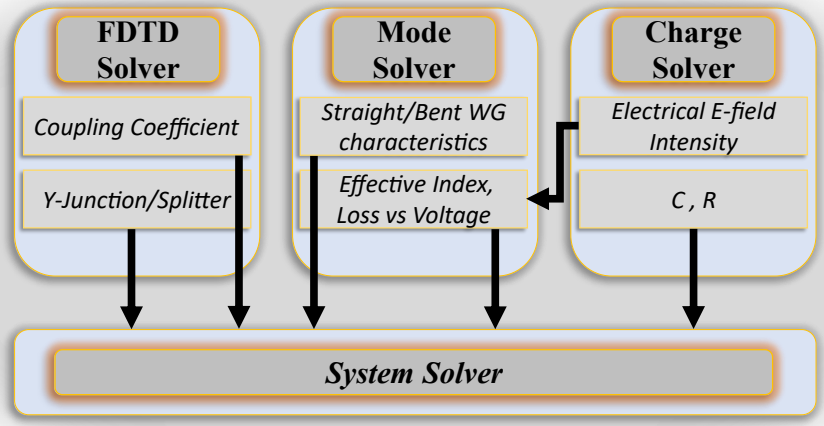

**Fig. 1 .**The graphical overview of the proposed building block-based simulation methodology.

An MRR can be represented as a series of subcomponents. An all-pass MRR uses a single-point directional coupler, a bent slotted waveguide with a length equal to the ring's circumference, and straight waveguides to form the bus waveguide (Fig. 2a). A similar approach can model add-drop MRRs by splitting the bent waveguide in half and adding another coupler. **Error! Reference source not found.**a and b depict the 3D scheme and the cross-sectional view of the studied all-pass SOH MRM. The dimensions of the studied SOH MRM include the microring radius ($R = 60\ \mu m$), coupling gap ($G_c = 0.55\ \mu m$), slab thickness ($T_s = 70\ nm$), total thickness ($T_r = 220\ nm$), rail width ($Wr = 230\ nm$), slot width ($Ws = 200\ nm$), and the waveguide to metal contact distance ($D = 8\ \mu m$). The polymer is coated on the MRM as cladding (Fig. 2b) [40]. Next, we provide step-by-step details of the simulation workflow for the SOH MRM.

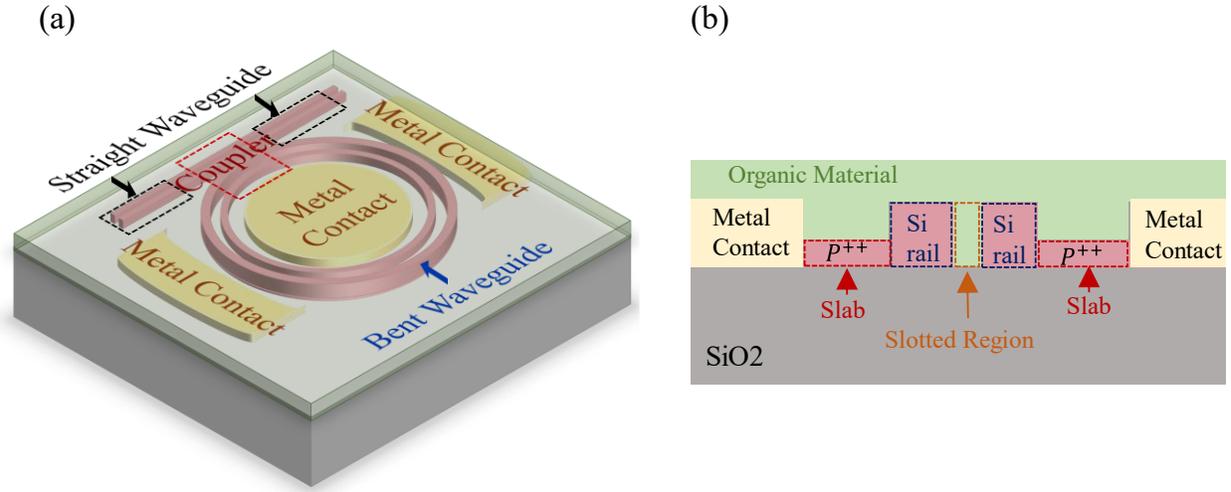

**Fig. 2**. The schematic of the investigated SOH MRR. **(a)** The 3D schematic, including the subcomponents, and **(b)** the cross-section view.

First, the optical characteristics of the coupler subcomponent are achieved using an FDTD simulator (e.g., Ansys/Lumerical FDTD Package). As the name suggests, the FDTD simulator is an optical simulation solver that solves Maxwell's equations in the desired geometry based on the FDTD method. It can find the optical frequency response, transmission, and reflection properties of the structure [54]. To find the coupling coefficient, a 3D FDTD simulation of the coupler section of the MRR (Fig. 2) is simulated in the FDTD solver. The organic material is defined as a material with a refractive index equal to 1.54 [40]. Since the coupling coefficient varies with the wavelength, the coupling coefficient of the fundamental TE modes in the wavelength range of 1.5-1.6 $\mu m$ is found (see Supplementary Fig. S1 online).

Next, the optical characteristics of the straight and bent slotted waveguides subcomponents are determined. Since the MRM structure is longitudinally uniform, a 2D simulation is conducted to obtain the optical characteristics of the bent and straight waveguides. The 3D simulation would significantly slow the workflow for incremental improvement in accuracy. Mode solver (e.g. MODE Lumerical package) calculates the optical modes by solving Maxwell's equations on a waveguide cross-section providing the mode field profile, effective index, group index, dispersion, and loss over a frequency range [54]. First, the modal analysis assesses the light confinement in the slotted straight and bent waveguides. The *x*-component of the optical E-field of the straight and bent waveguides at the wavelength of 1550 nm are demonstrated (see Supplementary Fig. S2. a and b online). Then, a frequency analysis is performed to find the key optical quantities of straight and bent waveguides in the desired wavelength range (1540-1570 nm) [40]. The real part of the effective index vs. wavelength in the range of 1540-1570 nm for the straight and bent waveguides can be found as Supplementary Fig. S2.c online. Since the loss calculated from the simulation is often significantly different from the experimental loss, the imaginary part of the effective index is not shown. This is mainly due to the sidewall roughness inside the slot, which is challenging to simulate accurately.

The next step is assembling all primitive subcomponents in a circuit-level simulator (e.g., Lumerical INTERCONNECT package) and loading their corresponding physics-level characteristics to attain the optical characteristics of the whole device. The Lumerical INTERCONNECT package is a photonic integrated circuit-level simulator that allows the co-simulation of optical and electro-optical effects in PICs [54]. Using the INTERCONNECT primitive library, the required elements corresponding to MRM subcomponents, including the waveguide coupler, mode waveguides, and optical Network Analyser (ONA), are placed and connected (see Supplementary Fig. S3). The grating couplers are selected from [54,61]. The mode characteristics of the waveguide exported from the MODE package is uploaded to the "mode waveguide" element. At this step, the waveguide edge roughness loss (35 dB/cm) [40] is added as the "excess loss" parameter for the bent and straight waveguides. The coupling coefficient data exported from FDTD solver, is also imported in the "waveguide coupler" element.

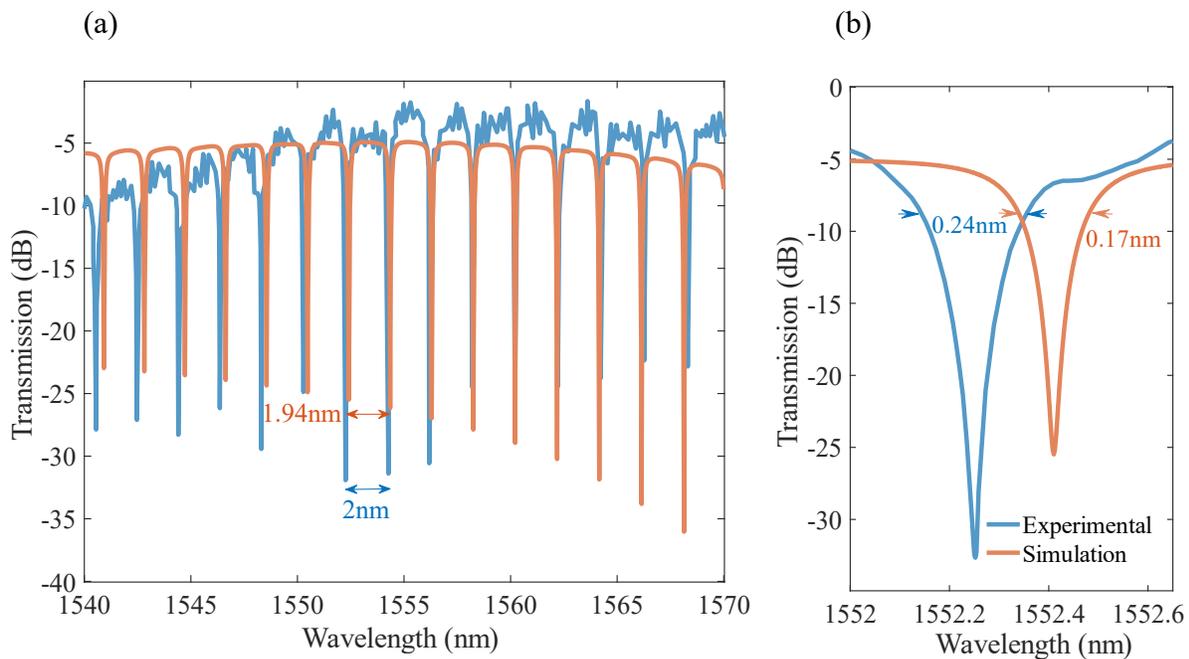

**Fig. 3** Modeling the optical transmission response of the MRM. (**a**) The transmission profile vs. wavelength showing the matched simulation and experimental results. Simulated FSR=1.94 nm is in line with the experimental results FSR=2 nm reported in ref [40]. (**b**) The transmission spectrum at 1552-1553 nm. The simulated 3dB bandwidth is equal to 0.17 nm, which is compatible with the experimental results (0.24 nm [40]). Neglecting experimental loss in simulation introduces a significant difference in ER and optical bandwidth, leading to an inaccurate simulation.

The ONA provides the optical characteristics of the SOH MRM, such as the transmission spectrum, FSR, Q-factor, FWHM, total loss, group delay, and group velocity. Fig. 3a shows the transmission profile of the SOH MRM in the wavelength range of 1540 to 1570 nm. To verify our simulation methodology, the experimental results have been presented [40]. The simulation results are in good agreement with the experimental data. The simulated FSR (1.94 nm) is close to the experimental measurements (2 nm) [40]. Fig. 3b illustrates the transmission spectrum in the 1552-1553 nm wavelength range. The simulation resonance wavelength (1552.41 nm) shows a negligible difference from the experimental one (1552.26 nm). The simulated 3dB bandwidth of 0.17 nm is also comparable to the experimental 3dB optical bandwidth of 0.24 nm [40]. The correlation between simulation and experimental transmission spectrum, specifically ER and Q-factor shows the significance of accounting for the experimental loss in simulation. The relatively small differences between the simulation results and experimental data can be attributed to differences in the dimension of the device, such as the slot width and coupling gap due to fabrication variations. For instance, a lower coupling gap ($G_c$) and larger slot width ($W_s$) lead to a higher coupling coefficient. As a result, this leads to a higher 3dB bandwidth and extinction ratio. The strip-to-slot mode convertor between the grating coupler and the straight bus waveguide is not included due to its low impact on the MRR transmission profile. Developing the simulation methodology based on Lumerical INTERCONNECT enables access to extensive optical elements of foundry PDK libraries resulting in more accurate simulation results.

## 2.2. Electro-optical active simulations

SOH modulators show a linear EO (Pockels) effect using highly nonlinear organic materials ($\chi^2$ nonlinearity) as cladding on the silicon waveguide [35]. The applied voltage to the metal contacts is dropped off through the $P^{++}$ doped slab waveguide to the silicon rails (Fig. 1**Error! Reference source not found.**b) and gives rise to an RF/DC E-field (electrical E-field) in the slotted region. The applied electrical E-field changes the refractive index of the organic material in the slotted section (equation (1)), where the optical field is highly confined. The optical nonlinear interaction of the confined light in the slotted region with the nonlinear organic material imposes a resonance shift which can be further utilized for optical signal modulations. The polymer's refractive index variation by the applied voltage is given by equation (1) [35,39,62,63].

$$\Delta n(E) = \frac{1}{2} \, r_{33} \, n_0^{\,3} E \tag{1}$$

where, $r_{33}$ and $n_0$ are EO coefficient and refractive index of the organic material, $E$ is the electrical E-field created by the applied voltage V to the modulator with slotted width defined as $W_s$.

To simulate the established electrical E-field in the slotted region of an SOH modulator at different voltages, a simulator that can account for free carriers and electrostatic potential is required. A charge simulator solving the Poisson's and drift-diffusion equations through the structure to calculate the electrostatic potential and free carriers' density profile ((e.g., Lumerical CHARGE solver). enables the simulation of active photonic devices through an exhaustive charge transport simulation [54]. At this step, a 2D charge simulation is done for the slotted sub-component of the proposed MRM. The silicon slotted waveguide, two metal contacts (100 nm Au on top of 1.5 $\mu m$ Al), and polymer (material with a relative static dielectric permittivity of $\varepsilon_r = 3$) clad are defined and placed. The metal contacts are located 8 μm from the slotted waveguide [40]. The background p-doping concentration equal to $5 \times 10^{+17} cm^{-3}$ and p++ slab waveguide doping level equal to $1 \times 10^{+20} cm^{-3}$ are introduced. The slope shape of the rail waveguides, which influences the electrical E-field value through the slotted part, is also considered in the simulation. Applying a voltage (1, 2, and 3 volts) to the metal contacts provides the electrical E-field intensity in and around the waveguide (Supplementary Fig.S4 a-c online). As shown, the electrical E-field is not uniform in the slotted waveguide and is not equal to zero at the outer region. In this simulation, a more accurate resonance shift is obtained by taking these non-uniformities into account, compared to the previous studies [44,53]. Previous studies take the electric field within the slotted region to be

$$E = \frac{V}{W_s},$$ (2)

and everywhere else to be zero, ignoring fringe effects. They ignore the effect of doping level in the slab waveguide, slab waveguide thickness, and side slope shape of the slotted section affecting $E$ amount. This results the effective index change vs. voltage 1.8 (for slot width 200nm) and 2.4 (for slot width 150nm, an inaccurate estimation of the slope side effect) times higher than the one achieved for the MRM [40] in our simulation methodology. Increasing the voltage increases the electrical E-field (see Supplementary Fig.S4 online). Increasing the doping level in the slab waveguide, decreasing the slot width, and increasing the rail waveguide slope due to fabrication, enhance the change in the electrical E-field vs. voltage. This further results in a higher refractive index change (Equation (1)). Our software-based simulation methodology facilitates the application of all effective factors to successfully design and optimize the SOH modulators. The theoretical calculation of SOH MRM performance by applying all the abovementioned effective parameters is highly complex. Furthermore, the building block approach in the adapted simulation technology allows a fast simulation of the electrical E-field at various voltages. The electrical bandwidth ($BW_{ele}$) of the SOH modulator ($\frac{1}{2\pi RC}$) calculated by the corresponding series RC circuit of the SOH slotted waveguide MRM [40] can be achieved in charge simulator. The resistance ($R$) of the SOH modulators is defined by the slab layer, depending on the slab thickness, width, length, and the doping's levels. The capacitance of the slotted SOH modulator is introduced by two silicon rails with dielectric organic material in between (semiconductor-insulator-semiconductor capacitor). $R$ is calculated by connecting one of the silicon rails to the same side metal contact to create a continuous circuit and finding the resistance using the current achieved from simulation and the applied voltage ($R = \frac{V}{I}$). The capacitance ($C$) is also calculated using the change in net charge over the voltage change. Considering half of the 80% circumference of the ring, $R = 0.3 \ \Omega.cm = 20 \ \Omega$ and $C = 0.09 \ fF/\mu m = 13.5 \ fF$ ($BW_{ele} = 580 \ GHz$) are achieved in simulation, which are comparable to $R = 0.15 \ \Omega.cm$ and $C = 0.07 fF/\mu m$ ($BW_{ele} = 1.5 \ THz$) reported in [40]. However, a higher experimental resistance (M $\Omega$) has been shown in [40] attributed to surface damages during the oxide-etching and improper doping, leading to a low experimental electrical $BW$ ($BW_{ele} = 1 \ GHz$). The MRM bandwidth is also constrained by the optical bandwidth as $BW_{opt} = \frac{w_0}{2\pi Q}$, where $w_0$ is the optical frequency and $Q$ is the MRM's quality factor ($Q = \frac{\lambda_0}{FWHM}$). Considering the simulations results in section2.1, the $BW_{opt} = 21 \ GHz$ is achieved, which is comparable to the $BW_{opt} = 30 \ GHz$[40]. The modulation bandwidth is limited by either the optical or the electrical bandwidth.

The subsequent step requires the effective index of the slotted waveguide at each voltage. The spatial distribution electrical E-field extracted from charge electrostatic simulator is imported to the Mode solver to obtain the essential waveguide optical characteristics at various voltages. In this step, the polymer material is defined as a new material in the MODE solver based on its EO effect characteristics described by equation (1), where $r_{33} = 19\,\frac{pm}{V}$ and $n_0 = 1.54$[40] and $E$ is the electrical E-field extracted from Charge simulator. The modal and frequency analysis are performed to achieve the effective index vs. voltage. The effective index vs. voltage data is imported to a system simulator (e.g., Lumerical INTERCONNECT package) to realize the modulation response. The simulation of the inserted electrical E-field profile from Lumerical CHARGE to the MODE FDE at 1V is shown as Supplementary Fig.S5a online. The effective index change as a function of voltage at the wavelength of 1552.41 nm has also been depicted as Supplementary Fig.S5 b online.

Next, the passive bent waveguide is replaced with the active one in the system simulator (e.g., Lumerical INTERCONNECT) to obtain the modulation response. From the INTERCONNECT primitive library, "DC source" and "optical modulator measured" elements are placed and connected in the current passive SOH MRM simulation platform (Supplementary Figure S3). The INTERCONNECT simulation platform of the active simulation of SOH MRM, including both optical and electrical stimuli has been depicted (see Supplementary Fig. S6 online). The "DC source" and "optical modulator measured" have been named "Bias Voltage_DC" and "Optical modulator", respectively. The former defines the electrical stimulus, and the latter refers to the active section of the SOH MRM.

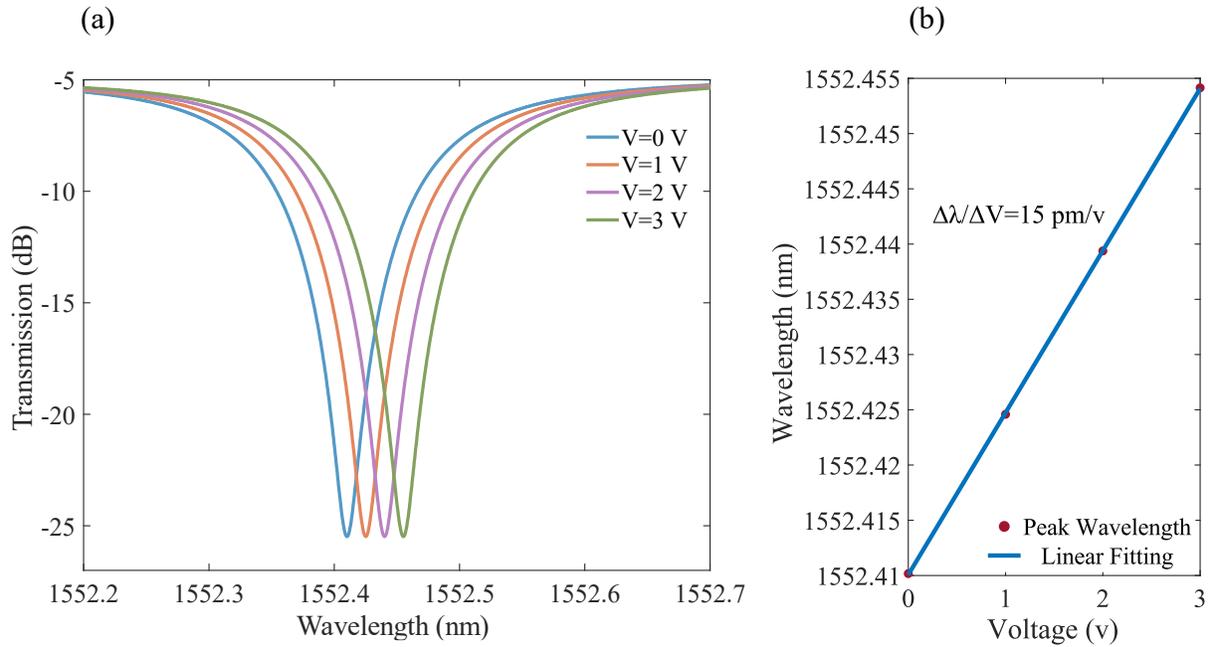

**Fig. 4** The electro-optical response of the MRM. **(a)** The voltage-induced resonance shift of the SOH MRR structure. **(b)** The peak resonance shift vs. voltage showing 15 pm/V SOH MRM modulation efficiency.

The INTERCONNECT circuit-level simulator achieves the resonance shift as a function of voltage ( Fig. 4a). Fig. 4b shows the linear fitting of the peak resonance shifts where the device tunability of 15 pm/v is comparable to the experimental value of 16 pm/V [40]. The negligible 1 pm/v difference in the resonance tuning can be attributed to the device dimension mismatch due to fabrication

variations—for instance, a higher slope increases both the optical and electrical E-field leading to a higher resonance shift. The developed building block-based SOH modulators simulation approach that connects device-level Lumerical packages to INTERCONNECT enables its compatibility with EDA-style full-flow PICs design methodologies from design to fabrication [11-18].

## 2.3. Topology Compatibility

The developed simulation methodology based on the building block approach is compatible with a wide range of SOH device topologies such as MZI [44], ring assisted MZI (RAMZI) [64], partially slotted MRMs [65], and any other complex combinations of them [66]. Any desired SOH modulator can easily be modeled by connecting the characterized active slotted waveguide with the required sub-components in the circuit-level simulator. The methodology also remains largely unchanged for travelling wave MZMs, with the addition of just an extra step: running a small signal simulation in CHARGE simulator to find the R and AC capacitance (real and imaginary parts of the impedance), transferring these parameters to the Mode solver for a mode simulation covering 1-100GHz frequencies to analyze RF properties of the transmission line (including RF effective/group index and RF impedance), and importing these results into the system simulator (INTERCONNECT) for calculating optical transmission and generating an eye diagram. Our methodology facilitates the design, characterization, optimization, and synthesis (phenomenological modeling) of various SOH modulators at no additional cost. To confirm the validity of the simulation methodology for any other SOH devices with any polymer materials, an experimentally demonstrated MZM [44] is simulated.

An MZI consists of one Y-splitter, one Y-junction, and two straight waveguides as phase shifters (PS). Since the studied MZM is based on the slotted waveguides, two strip-to-slot mode convertor sub-components are also considered. Fig. 5 shows the 3D schematic of the studied slotted MZM comprised of the building blocks. The cross-sectional view is analogous to the one shown in Fig. 1b. The dimensions of the SOH MZM include the silicon slab thickness $T_s = 70 nm$, waveguide thickness $T_t = 220 \, nm$, rail width $w_r = 70 \, nm$, and slot width $W_s = 160 \, nm$, and length $L = 1.1 \, mm$ [44].

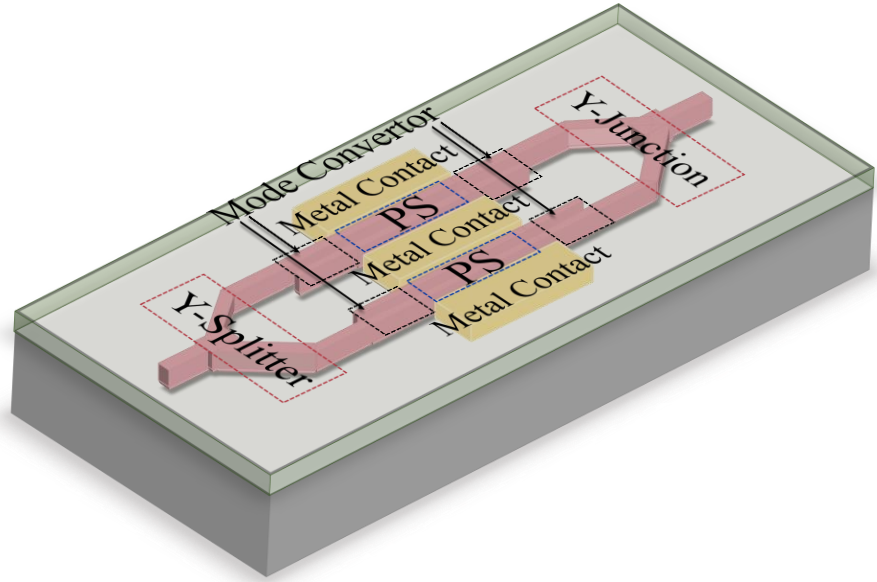

**Fig. 5.** The 3D schematic representation of a slotted MZM accompanied with its building blocks.

The simulation steps of the SOH MZM are the same as those outlined in the MRM example in earlier sections. However, the FDTD simulation to find the coupling coefficient in an MRR and the mode simulation for bent waveguide characterization are not required. The optical E-field at the wavelength of 1550 nm, the electrical E-field at 0.9 V, and the electrical E-field profile imported in MODE of the MZM can be found as Supplementary Fig. S7 a, b, and c, respectively. Cascading the characterized slotted straight waveguide with two Y-junction/splitter components and mode-convertors [67] (that exist in the interconnect or PDK library) in the INTERCONNECT Lumerical package (see Supplementary Fig. S8) results in the SOH MZM transmission profile (Fig. 6). According to the transmission profile, applying 0.8 V leads to a $\pi$ phase shift ($V_\pi = 0.8\ V$), which is comparable to the experimental results ($V_\pi = 0.9$ V [44]). Therefore, our precise simulation approach can attain well-matched experimental results. The 0.1 V margin in experiment and simulation results can arise from several reasons, such as a mismatch in doping level, contact-waveguide distance, and dimensional fabrication variations. For instance, a larger distance between metal contacts and waveguide, lower slab layer doping level, and having no background doping can cause lower electrical E-field intensity, which results in lower resonance shift and higher $V_\pi$.

## 3. System-level simulations

Our building block-based simulation methodology is well-suited for accurately modeling SOH-based PICs. Here, a compact model of the designed SOH modulators is generated to further facilitate the scalability and integrability with other passive and active photonic components and electronic devices. Integrating all the characterized subcomponents in INTERCONNECT and selecting the "create a compact model" object generates the SOH modulator compact model. The generated compact model is added to the INTERCONNECT library for further system-level simulations. All steps from the physics-level to the system-level are derived automatically using the workflow object in the CHARGE solver. Ansys/Lumerical CML Compiler also automatically produces the compact model based on the data extracted from physics simulation solvers. This eases the modeling of SOH modulators for photonics designers.

## 3.1 Scalability & Integratibility

As we have shown, our novel simulation approach outweighs the studied methodologies in many aspects. The two key advantages of the proposed simulation methodology are easy and fast scalability and integrability. This is particularly useful for enabling new applications (in sensing, microwave, quantum, and neuromorphic computing) that require fast and accurate simulation of large-scale PICs with increasing complexity in terms of material, device, and circuit integration from hybrid to heterogenous to monolithic integration. From a system-level perspective, the proposed building block-based modeling approach integrate with a circuit simulator (e.g., INTERCONNECT), supports foundry-provided PDK library elements, and further offers benefits in calculating PICs behavior accurately.

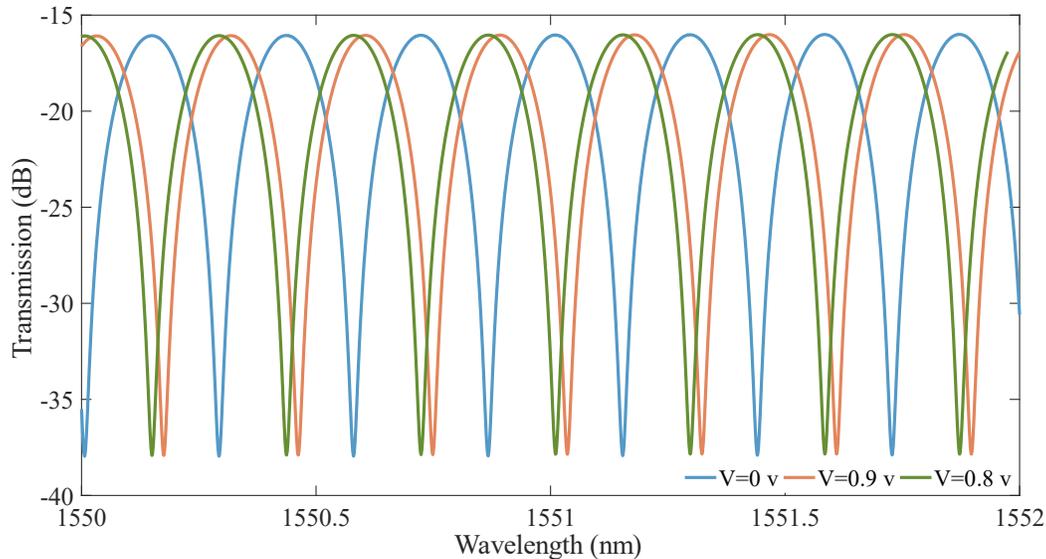

**Fig. 6.** Optical and electro-optical response of the MZM. The transmission profile vs. wavelength at V=0, V=0.9 and V=0.8.

For instance, WDM (de)multiplexer, as a key part in PICs, is used in different applications, including optical communication [68], optical interconnects in datacenters [69], weightbank in neuromorphic photonics [7], and as a linear front-end in optical signal processing in RF photonics [70]. As a showcase example, we design, model, and study a 3-channel SOH MRM-based WDM (de)multiplexer built on the studied SOH MRM in sections 2.1 and 2.2, to represent the easy and fast scalability and integrability of the developed simulation methodology for PICs. The SOH-based WDM (de)multiplexer show higher modulation speed and lower power consumption compared to silicon WDMs due to high speed and low power SOH modulators [35]. The MRMs radii are designed as 60, 60.023, and 60.046 $\mu m$ to illustrate a low optical crosstalk. A SOH MRM with smaller radius enables higher channel WMD (de)multiplexer. Fig. 7a demonstrates the developed system-level optoelectrical co-simulation platform showing the proposed WDM system, using the SOH MRM compact model. The transmission profile of the 3-channel WDM spectrum at the wavelength range of 1540-1570 nm is shown in Fig. 7b. The resonance wavelengths equal 1552.41, 1553.08, and 1553.75 nm ( Fig. 7c). The eye diagrams of channels one, two and three for 10Gb/s signaling with a drive voltage of 3V and 1024 samples per bit are shown in Fig. 7.d, e, and f, respectively. For simplicity, we have used optical band pass filters at the receiver part. The proposed structure has also the potential of being served as a tunable and low power multi-channel filter or switch. Designing smaller radii SOH MRMs or partially slotted SOH racetrack modulators [65] enables a higher number of channels and higher speed. In principle, SOH MRMs with smaller radii will lead to higher bandwidth WDM systems.

(a)

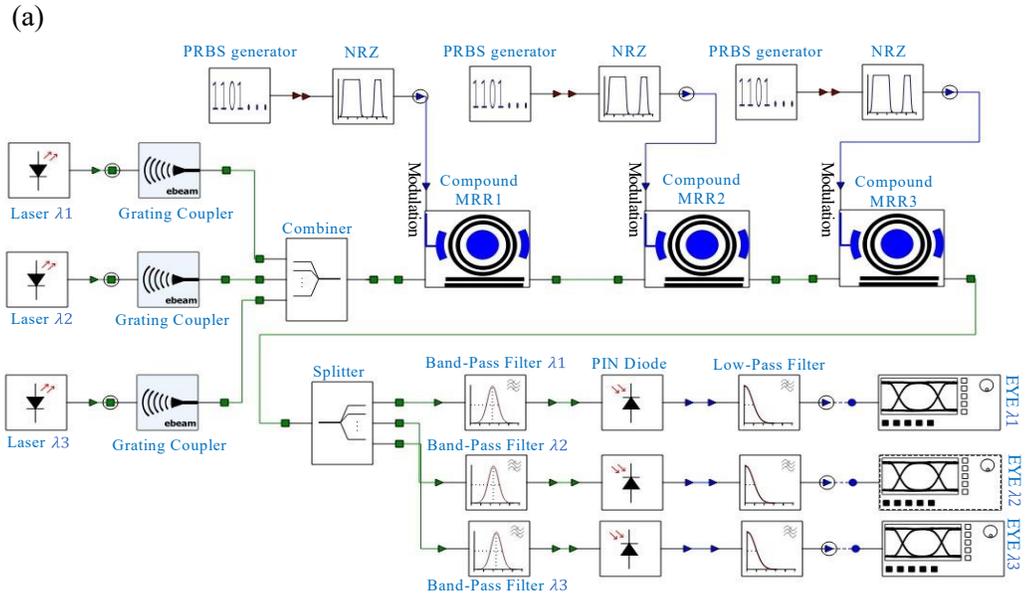

(b)

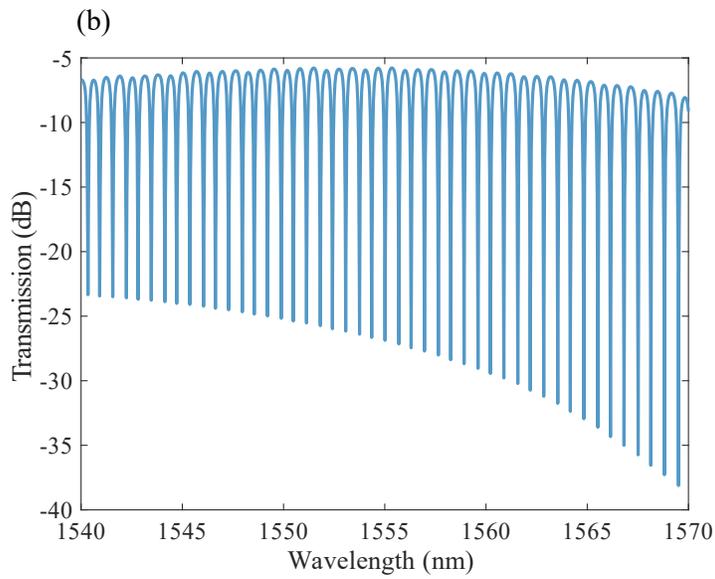

(c)

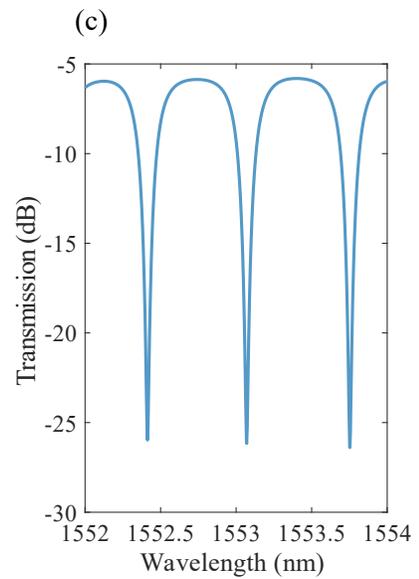

(d)

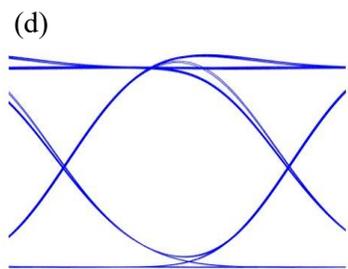

$BER = 0, Q - factor = 21$

$ER = 5.34dB$

(e)

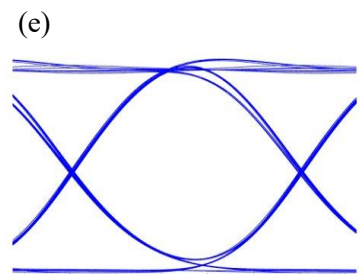

$BER = 0, Q - factor = 24$

$ER = 5.92dB$

(f)

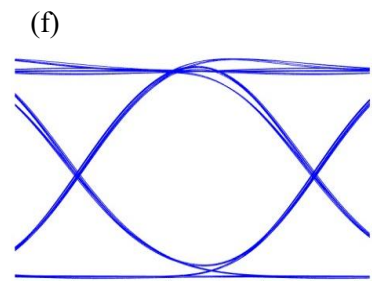

$BER = 0, Q - factor = 23$

$ER = 5.84dB$

**Fig. 7.** Three-channel WDM (de) multiplexer modeling. **(a)** The simulation platform of the proposed SOH MRMs WDM PIC. The radii are 60, 60.023 and 60.046 $\mu m$. **(b)** The corresponding transmission profile at 1540-1570 nm and **(c)** at 1552-1554 nm. **(d)** The eye-diagram of channel 1, **(e)** channel 2, and **(f)** channel3 at 10Gb/s and 1024 samples per bit.

## Conclusion

We have developed an accurate and comprehensive building block-based, phenomenological simulation methodology, spanning from physics to the system level, enabling the modeling of various SOH modulator topologies, including MRM, MZM, and ring assistant MZM. Our simulation methodology is compatible with EDA-style silicon PICs modeling approaches. This facilitates seamless and fast, modeling of large-scale SOH modulators integrated with silicon PICs, addressing a challenge in current SOH modulator modeling methodologies. The current theoretical and numerical modeling methodologies for SOH modulators are limited by computational resources for modeling large-scale SOH modulators. Furthermore, these methodologies lack the capability to integrate SOH modulators with silicon PICs and model the entire system within a unified platform. We are tackling these challenges by developing compact models for SOH modulators, allowing for the modeling of large-scale SOH modulator systems and their seamless integration with silicon PICs. As a representative example, we have demonstrated a 3-channel SOH MRM WDM (de)multiplexer—a commonly employed PIC in diverse applications, including communications, data center interconnects, neuromorphic photonics, sensing, and switching networks. Our methodology is compatible with any photonics-electronics co-simulation software, such as Ansys Lumerical and Cadence Virtuoso—two well-established and commercialized photonics-electronic co-simulation platforms for modeling large-scale PICs. Our developed simulation methodology can be readily applied to similar materials exhibiting Pockels and Kerr electro-optic effects, such as lithium niobate (LiNbO3) and BaTiO2, enabling the modeling of LiNbO3 and BaTiO2 modulators and their large-scale integration with silicon PICs. LiNbO3 and BaTiO2 modulators suggest high-speed, low-loss, and high modulation efficiency, leading to large-scale, high-speed, low-loss, and energy-efficient PICs.

## Funding